# From x-ray telescopes to neutron scattering: using axisymmetric mirrors to focus a neutron beam

B. Khaykovich[1*], M. V. Gubarev[2], Y. Bagdasarova[3], B. D. Ramsey[2], and D. E. Moncton[1,3]

[1]Nuclear Reactor Laboratory, Massachusetts Institute of Technology, 138 Albany St., Cambridge, MA 02139, USA

[2]Marshall Space Flight Center, NASA, VP62, Huntsville, AL 35812, USA

[3]Department of Physics, Massachusetts Institute of Technology, 77 Massachusetts Ave., Cambridge, MA 02139, USA

**Abstract**

We demonstrate neutron beam focusing by axisymmetric mirror systems based on a pair of mirrors consisting of a confocal ellipsoid and hyperboloid. Such a system, known as a Wolter mirror configuration, is commonly used in x-ray telescopes. The axisymmetric Wolter geometry allows nesting of several mirror pairs to increase collection efficiency. We have implemented a system containing four nested Ni mirror pairs, which was tested by focusing a polychromatic neutron beam at the MIT Reactor. In addition, we have carried out extensive ray-tracing simulations of the mirrors and their performance in different situations. The major advantages of the Wolter mirrors are nesting for large angular collection, and aberration-free performance. We discuss how these advantages can be utilized to benefit various neutron scattering methods, such as imaging, SANS, and time-of-flight spectroscopy.

# 1. Introduction

Neutron scattering methods are among the most important tools for studying the structure and dynamics of matter on a wide range of scales from atomic through mesoscopic to macroscopic. The powerful new Spallation Neutron Source (SNS) at Oak Ridge National Laboratory is stimulating new discoveries and creating novel opportunities for research using neutrons. Nevertheless, neutron methods will remain limited, sometimes severely, by available neutron fluxes. The efficient use of existing sources can therefore be as important a path toward more powerful instruments as the development of brighter sources. Consequently, much effort is spent to develop neutron guides and focusing optics that preserve neutron source brilliance [1-5]. Nevertheless, the tool-box of neutron instrument components is still relatively small, especially compared to that of x-rays, where much brighter and smaller sources are available and the dawn of x-ray lasers is in sight. New techniques for manipulating neutron beams might bring significant, even transformative, improvements of neutron instrumentation, and enable new science. In this paper, we demonstrate such a new technique: grazing-incidence mirrors based on full figures of revolution. In 1952 Hans Wolter described the advantages of glancing-angle mirrors for an x-ray microscope based on the pairing of a confocal ellipsoid and hyperboloid [6]. Wolter optics can be used as a lens, in order to either focus a beam or form an image of an object; such optics is especially useful for imaging, small-angle neutron scattering (SANS) and other neutron techniques.

Two major problems exist with grazing incidence optics. First, collection efficiency is small because of the small critical angle. (The collection area of grazing incidence mirrors is the projection of the mirror surface onto the aperture plane.) Second, grazing incidence mirrors suffer from large aberrations, distorting the image in the focal plane. Wolter mirrors are capable of improving performance in both areas. Nesting multiple mirrors increases collection area by placing co-axial mirrors one inside another, as routinely done in x-ray astronomy. And a multilayer coating will further increase the critical angle and thus collection efficiency. For aberrations, the strongest ones are removed by meeting the Abbe sine condition, which requires that all geometrical paths through the optical system produce the same magnification. With grazing incidence optics, the neutrons must undergo two reflections between source and image, which is exactly what occurs in the Wolter design. Eliminating aberrations means that a true image of an object can be formed, and this is important well beyond imaging methods, such as for SANS. Also, all the reflected neutrons travel the same distance from source to image, a property important for some time-of-flight applications. (Technically, the path length is the same only for neutrons hitting one mirror pair, but in nested configurations the differences are very small). Maximizing collection efficiency of the mirrors while eliminating distortions will lead to novel neutron instruments.

In this paper we demonstrate focusing of a polychromatic neutron beam by four nested nickel Wolter mirrors at the MIT Nuclear Reactor. For the proof-of principle experiment we used the mirrors to de-magnify a 2 mm diameter source by a factor of 4; both the source and detector were placed at focal planes of the optics, 3.2 m apart. We found that the performance of the mirrors is in good agreement with expectations based on geometrical optics. In addition to experimental measurements, ray-tracing simulations were used to analyze the performance of the mirrors for collecting neutron flux. As an example, we simulated a system with source-sample separation of 10 m. It was found that the mirrors would increase the neutron flux on a sample of 1 mm diameter by almost an order of magnitude, for 5 meV neutrons and mirrors coated with high-critical-angle multilayers (m=3 neutron supermirror coating). Beyond simple flux concentration, Wolter mirrors offer unique advantages, which make them suitable for a number of neutron applications, as we discuss below.

## 2. Geometrical optics of Wolter mirrors

Figure 1(a) shows the geometry of the Wolter type-I mirrors. Incident rays must reflect from both mirrors before coming to a focus. Since only double-reflected rays make it to the focal plane, the rays which will not intersect the first mirror are stopped by a beam stop in front of the mirrors. The system is defined by the following initial parameters: the radius at the intersection of two mirrors, the distance from the source focal plane to the intersection, grazing angles at the intersection, and the length of the mirrors. Finally, the mirrors are made such that the grazing angle for each mirror is the same close to the intersection point, and so are the lengths of the mirrors [7]. This condition ensures that the part of the beam reflected from the first mirror is intersected in full by the second mirror.

Figure 1(b) shows the geometry of the mirrors close to the intersection point. The conditions for the angles on Figure 1 are as follows:

$$\tan\theta_1 = r_i/f_s, \quad \tan\theta_2 = r_i/f_i \tag{1}$$

$$\phi_E = (\theta_2 - 3\theta_1)/4, \quad \phi_H = (3\theta_2 - \theta_1)/4, \tag{2}$$

$$\theta = \theta_1 + \phi_E = \theta_2 - \phi_H \tag{3}$$

The ellipsoid and hyperboloid are defined respectively by the equations:

$$r_E = b_E\sqrt{1-(z-z_{0E})^2/a_E^2}, \quad r_H = b_H\sqrt{(z-z_{0H})^2/a_H^2 - 1} \tag{4}$$

Here $z$ is the coordinate along the optical (beam propagation) axis, $x$ and $y$ are perpendicular to $z$, and $r^2 = x^2 + y^2$. The parameters $a$ and $b$ denote the semi-major and semi-minor axes of the ellipsoid and hyperboloid, while $z_0$ denotes the location of their centers. From the initial parameters and the confocality condition for the hyperboloid and ellipsoid mirrors, we derived mathematical expressions for $z_{0H}$ and $z_{0E}$:

$$z_{0E} = c_E, \quad z_{0H} = 2c_E - c_H = 0.5(f_s + f_i) + z_{0E}, \quad c_H = \frac{f_i\theta}{r_i/f_i - 2\theta}, \tag{5}$$

Parameters $a$ and $b$ are found using the definitions for ellipsoid and hyperboloid: $a_E^2 - b_E^2 = c_E^2$ and $a_H^2 + b_H^2 = c_H^2$ and by using the equations for ellipsoid and hyperboloid at the intersection point $r_E = r_H = r_i$.

The intersection radius increases with $\theta$. Therefore, mirrors made of high critical angle material can be nested around low-critical-angle ones. For example, both $Ni^{58}$ and neutron supermirror multilayer coatings have larger critical angles than that of Ni for the same energies. Hence, $Ni^{58}$ and multilayer coated mirrors can be nested around a small Ni mirror. Obviously, mirrors with the same critical angle can be nested too.

## 3. Experimental results

A system of four nested ellipsoid-hyperboloid Ni mirror pairs was made and tested at the MIT Reactor. Our Wolter mirrors are made by a replication technique developed for the x-ray astronomy [8,9] and x-ray imaging applications [10]. A mandrel of desired geometry is coated with nickel (or nickel-cobalt alloy) in an electrochemical bath. When a Ni shell reaches desired thickness, it is separated from the mandrel. The mirrors described here were made using existing mandrels [10], which were appropriate for using at the Neutron Optics Test Station at the MIT Reactor. Numerical parameters for each of the four mirror pairs are listed in Table 1. Each row in Table 1 corresponds to one of the four nested mirror pairs. The diameter of each mirror is such that it does not project a shadow onto a larger mirror. The optics have a magnification $M = 1/4$; when the angles between rays and the optical axis are small (paraxial approximation in geometrical optics), angular and lateral magnifications are equal to each other, $M = \theta_1/\theta_2 = f_i/f_s$. The origin ($Z = 0$, $r = 0$) is at one focus of the ellipsoid. The source is at the origin, and the detector is at $Z = 3.2$ m. The focal distances are: $f_i = 640$ mm, $f_s = 2560$ mm. The projected length of the hyperbolic section along the optical axis is $L_H$, and that of the elliptical section is $L_E$. The grazing angle for a ray from the origin to the intersection point is also reported in Table 1. Hyperbolic and elliptical shapes are given by equations (1) – (5).

**Table 1**. Optical Parameters

| $a_H$ [mm] | $b_H$ [mm] | $a_E$ [mm] | $b_E$ [mm] | $L_H$ [mm] | $L_E$ [mm] | $r_i$ [mm] | $\theta$ [deg] |
|---|---|---|---|---|---|---|---|
| 533.2821 | 7.296319 | 2133.382 | 14.59266 | 30.000 | 31.097 | 14.298 | 0.40000 |
| 533.2827 | 7.665439 | 2133.393 | 15.33097 | 30.000 | 31.097 | 15.021 | 0.42022 |
| 533.2824 | 8.053217 | 2133.404 | 16.10662 | 30.000 | 31.097 | 15.781 | 0.44148 |
| 533.2811 | 8.460593 | 2133.415 | 16.92151 | 30.000 | 31.096 | 16.579 | 0.46381 |

The mirrors were placed in the polychromatic thermal neutron beam. Figure 2 shows the photograph of the mirrors installed at the beam-line. A 2 mm diameter cadmium aperture source and a detector were positioned in two focal planes. The detector is based on a standard neutron-sensitive scintillator screen (Li-doped ZnS). The light output from the screen is detected by a CCD (Andor Luca EMCCD). The spatial resolution of the detector was calibrated by imaging a 1.2 mm pinhole in a Gd foil. By fitting the image with a Gaussian, we found that the pixel size was (92 ± 4) μm, FWHM.

The image of the neutron source demagnified by the mirrors is shown on Figure 3. Half-power diameter (HPD) of the spot is 0.62 mm. We scanned the detector along the beam axis and measured the size, HPD, of the resulting image. The HPD as a function of the distance from the nominal focal plane is plotted on Figure 4, together with the values calculated by ray-tracing. Figure 5 shows the result of an angle scan, when the mirrors were tilted with respect to the beam axis in horizontal plane. Finally, Figure 6 shows the results of an experiment when the detector was placed between the optics and the image focal plane. In these measurements, the optics had only two nested mirror pairs. Figure 6a shows a ring formed by the reflected neutrons, when the detector is close to the focus. When the detector is further away from the focus, as on Figure 6(b), two rings can be resolved, each formed by reflections from one of the two mirror pairs. These results confirm that both mirror pairs contribute to deflecting the beam towards the focal point. Ray-tracing calculations produce the rings of similar size, as shown on Figure 6(c)**.** Further comparison between the experiment and simulations are detailed in Section 5 below.

The mirrors described above were optimized for the ease of testing and manufacturing. Since they are relatively short and made of Ni, the mirrors are not optimized for flux collection. For example, ray-tracing calculations predicted that only neutrons of up to about 5 meV are focused. These cold neutrons constitute a small fraction, about 5%, of the thermal neutron flux at the MIT Reactor. Supermirror multilayer coating will increase the upper cut-off energy, and therefore the collection efficiency of the mirrors. Also, longer mirrors will collect higher portion of the neutron flux. We modeled flux collection efficiency of supermirror-coated long Wolter optics by ray-tracing simulations, as described in the next section.

## 4. Neutron flux collection by Wolter mirrors: ray-tracing simulations

Here we show the results of ray-tracing simulations, which could be suitable for a neutron instrument installed at SNS or other large neutron facilities. A standard software package, McStas, was used for the simulations [11,12]. Each mirror was included as a separate McStas component. Confocal Wolter pairs were combined into a McStas instrument, which included a neutron source, monitors and other necessary beam-line components. The ray-tracing algorithm used for the mirror components is similar to that used for neutron guides. McStas supplies and tracks coordinates and velocities of the neutrons, $\mathbf{R} = (x,y,z)$ and $\mathbf{v} = (v_x, v_y, v_z)$. The neutron beam first propagates from the source to the opening aperture plane of the mirrors. Neutrons falling outside the leading edge of the mirror are discarded. Next, the position of the intersection of the neutron trajectory with the mirror is determined, $\mathbf{r_p} = (x_p, y_p, z_p)$. Since the mirrors have finite lengths, only a fraction of trajectories intersect the mirrors. For the reflected neutrons, $\mathbf{k'} = \mathbf{k} + 2(\mathbf{k}\cdot\mathbf{n})\mathbf{n}$. Here $\mathbf{k}$ and $\mathbf{k'}$ are wave-vectors before and after the reflection and $\mathbf{n}$ is the normal to the mirror at $\mathbf{r_p}$, as calculated from the mirror geometry. The reflectivity is a function of $\mathbf{Q} = \mathbf{k'}-\mathbf{k}$ exactly as for flat mirrors. Reflected neutrons receive a weight proportional to the reflectivity. (McStas uses the "weight factor" to calculate how many neutrons are transmitted through an instrument or a component. For example, if the reflectivity of a mirror is 10%, then every neutron is reflected, but assigned the weight of 0.1.) Reflected neutrons, with the new momenta $\mathbf{k'}$, propagate further to the entrance of the next component.

We calculated the neutron flux density collected by the optics and then optimized the system's parameters for maximum flux at the sample. Mirrors with the following characteristics were used as an example: source-to-sample distances of 10 and 25 m and critical angle of 21 mrad (corresponding to an m = 3 supermirror multilayer coating and 5 meV neutrons). Several magnifications between 0.1 and 1 were tested. The flux was maximized by changing the mirror intersection radius $r_i$. Considering current manufacturing constraints, the length of each mirror was limited to <10 $r_i$ and < 0.7 m. We found that the maximal flux density at the sample was about 3 times that of the source for one mirror pair, as shown on Figure 7. At first, the collected flux increases with the radius, but then the flux starts to decrease. The decrease starts when some of the neutrons begin to intersect the second, hyperboloid, mirror with the angle above the critical angle. In our geometry, maximum flux is achieved at magnification 0.1 for both systems, of 10 m and 25 m as shown on Figure 8. Figure 9 shows the effect of nesting on collection efficiency of Wolter optics. Increasing the number of nested mirrors leads to a significant increase in the neutron flux on the sample. A system of 4 nested mirrors produces about 8 times the flux density of the source. The flux density does not depend on the size of the source for different magnifications and source radius between 1 and 10 mm, according to ray-tracing simulations. The independence of flux on the source size is the consequence of low aberrations for off-axis rays. The ability to change the source size without affecting the performance of the optics is important for many applications, since the samples often come in various sizes.

## 5. Discussion

The experimental measurements were made using small Ni mirrors, and the results are compared to calculations. In addition, extensive ray-tracing modeling was done for long m=3 supermirror-coated optics. This Section is sub-divided accordingly.

### A. Experimental results

The experimental results shown on Figures 3 and 4 were compared with ray-tracing simulations, using parameters from Table 1. The neutron flux is nearly constant across the 2 mm diameter source. Therefore, if the mirrors are perfectly aligned and have small manufacturing errors, we expect the focal spot to be 0.5 mm diameter—4 times smaller than the source. The corresponding HPD is 0.35 mm, as confirmed by ray-tracing simulations shown on Figure 4. The HPD of the focal spot was measured to be 0.62 mm. The mirrors are manufactured with small deformations (usually called figure errors), but imprecise machining of the holder resulted in unusually large figure errors during our measurements. (A small distortion of the focal spot is actually visible on Figure 3.) Such deformations will be avoided in the future by adjusting the size of the holder.

Figure 4 shows how HPD changes with distance from the focal plane, both in experiment and simulations. By moving the detector along the beam-propagation axis, we measured the depth of focus, which is the extent of the region around the image plane in which the beam is focused into a spot with maximum intensity at the center. Away from the focal plane, the focal spot is transformed into rings shown on Figure 6. In the experiment, the depth of focus is about 40 mm. In simulations, it is less than 10 mm. Both the focal spot size and the depth of focus are affected by the mechanical deformation of the mirrors caused by improper size of the mirror holder.

We measured the flux density at the focal spot position with and without the mirrors, as deduced from the intensity registered by the CCD pixels. The experimental ratio is ($3 \pm 0.5$), whereas the ray-tracing simulations predicted the ratio of ($8 \pm 1$). The discrepancy is explained by taking into account the difference in area of the focal spot. The HPD in the experiment is almost twice that in simulations (the area of the focal spot is 4 times larger) and therefore the density is 4 times smaller than predicted. An additional factor contributing to the discrepancy in flux measurements is the energy spectrum of the thermal neutron beam. Since only neutrons below 5 meV are focused, an accurate flux comparison must include the neutron beam spectrum. We modeled the neutron flux spectrum of the MIT reactor using a Maxwell-Boltzmann distribution, and we found that the flux density is practically insensitive to the details of the source spectrum within our resolution. Finally, the detection efficiency of the scintillator might be energy dependent, but we believe this is not significant. We conclude therefore that the measurements are consistent with the ray-tracing model, and they raise the basic question of how to further increase collection efficiency.

Figure 7 shows that the neutron flux density at the focal point increases with $r_i$ until the grazing angle reaches the critical value. Therefore, larger critical angle allows larger diameter optics, which has larger collection efficiency. Consequently, for lower-energy neutrons used in many neutron applications, the flux density ratio will be larger than 10, for mirrors such as those in our examples. Let us compare simulation results with the fundamental limit of concentration efficiency. The theoretical limit of concentration is understood as follows. For simplicity, assume a circular source at infinity subtending a semi-angle $\theta_1$. Further assume the irradiation is concentrated onto a sample, subtending a semi-angle $\theta_2$. In this case, the limit of the concentration ratio is $C_{max} = \sin^2\theta_2/\sin^2\theta_1 = 1/M^2$, where M is the magnification of the optics [13]. If, for example, M = 0.1, maximal possible concentration $C_{max} =$

100. Therefore, for our theoretical example shown on Figure 9, the concentration of 8 is about 12 times smaller than the maximum possible concentration. Clearly, the losses are due to the cross-section of the mirrors in the beam. The cross-section can be increased if higher critical angles can be used, either by using longer-wavelength neutrons or multilayer coatings. Of course the choices will be determined by the specific applications.

**B. Wolter mirror applications**

The optical design of grazing-incidence focusing mirrors for neutrons is similar to that for x-rays, because the values of critical angles are similar for neutron supermirrors (for thermal neutrons) and dense x-ray mirrors (for few keV x-rays). On the other hand, some fundamental characteristics of the neutron and x-ray sources are quite different. Synchrotron x-ray sources provide well-collimated, small and extremely bright beams with angular divergence of 0.01 mrad and source size of 10 to 100 μm; x-ray sources studied by telescopes have angular divergences of 1 μrad. In comparison, the neutron sources are relatively large and diffuse with beam divergences of 0.1 to 1 degrees and beam source sizes measured in centimeters. Although the requirements for the shape and size of neutron focusing mirrors will be quite different from those for x-ray mirrors, the great success of synchrotron x-ray optics and x-ray telescopes provides both inspiration and experience for developing neutron focusing optics [2,5,14-17]. We believe that the neutron users' community can benefit enormously from the state-of-the-art technologies developed for x-ray optics.

Currently, several kinds of neutron focusing optics exist. Elliptical Kirkpatrick-Baez (KB) mirrors have been recently demonstrated following their successful use for focusing synchrotron x-rays [2,3]. The KB mirrors focus small beams, while nearly preserving source brilliance. The mirrors can be precisely figured with low roughness by mechanical bending of commercial neutron supermirrors. KB mirrors work best for sources of less than 1 mm, but they are not ideal for large neutron sources of 5 – 50 mm, as the length becomes prohibitive. In contrast, Wolter mirrors can collect neutrons from larger sources, especially if several nested mirrors are used. Furthemore, two reflections in Wolter geometry leads to collecting larger beam divergence than that possible with KB mirrors. Finally, elliptical KB mirrors are not ideal as imaging devices, since the magnification of an elliptical mirror is different for every point on the surface, leading to distortions in imaging of large objects. Wolter mirrors are constructed to minimize aberrations. Another example of focusing neutron optics is the elliptical neutron guide, which has been recently demonstrated [1]. Long elliptical guides focus the beam between neutron sources and samples, while the beam is transported over large distances. Wolter mirrors may serve a similar purpose, but they can be installed in such a way as to be easily removable if flexibility in focusing is required. Nesting allows the mirrors to be shorter than guides to achieve similar level of flux concentration.

The concentration ratio of various types of reflecting focusing optics depends fundamentally on the length and critical angle of reflecting surfaces. The requirements of specific neutron instruments for length, angular divergence, spatial and temporal resolution, cost, etc. will ultimately dictate the preferred focusing method. We believe the availability of a flexible focusing technique based on Wolter mirrors will enable improved instrument designs and lead to new science. We consider a few such improved instruments below.

SANS instruments often use focusing devices, such as lenses or collimators [18]. Lenses are not well suited for most time-of-flight (TOF) instruments because of chromatic aberrations. (The focal distance of biconcave neutron lenses changes as the second power of the neutron wavelength, $f \sim 1/\lambda^2$.) In fact, chromatic aberrations reduce the resolution even on reactor-based SANS instruments since the beam is not perfectly monochromatic. (Recent developments of magnetic lenses show some promise to reduce the chromatic aberration problem by modulating magnetic field. But such devices are very complicated and the need for polarized neutrons reduces count rate [19,20].) Focusing mirrors are free of chromatic aberrations. Therefore, a focusing toroidal mirror is being used in the Julich SANS instrument at FRM-II, but with limited success. This instrument, which is equipped with a Cu-coated mirror, requires large samples to produce enough signal [17]. In addition to the low critical angle of Cu coating, the length of the mirror is limited to 1.2 m by significant distortions (aberrations) in the focal plane [17]. Wolter optics offer significant improvements compared to toroidal mirrors. The use of an advanced coating combined with nesting and full figures of revolution should lead to drastic enhancement of the flux on the sample compared to the existing facility.

For absorption imaging, Wolter mirrors can play the same role as lenses in optical microscopes, and therefore can lead to dramatic improvements in the spatial resolution of imaging instruments. One possible design is a neutron microscope equipped with a Wolter optics lens, which creates a magnified image of the sample at the detector. If Wolter mirrors are placed between the sample and the detector, the resulting image will be magnified by a factor of 10 or more, using current technology. (Refractive lenses can be used for neutron imaging [21]. However, strong chromatic aberrations preclude their use in modern neutron imaging instruments.)

Since neutrons are only weakly absorbed in most materials, phase-contrast imaging promises substantially better image quality over absorption methods [22,23]. In the phase-contrast method, variations in the index of refraction are mapped, as opposed to variations in absorption. Phase-contrast imaging involves illuminating an object by a partially-coherent neutron beam, generally obtained at a large distance after transmission through a pin-hole (Fraunhofer diffraction regime). The image is obtained when a detector is placed at a distance from the sample (in contrast to absorption imaging, when the detector is right behind the sample). The combination of a small pin-hole source and large distances results in a weak signal. The signal could improve if Wolter mirrors were placed behind the sample, such that the image is focused on the detector. The phase coherence is preserved due to nearly aplanatic design of the optics. All the neutrons travel the same distance and so the relative phase is preserved if one pair of mirrors is used, and nearly preserved for a nested system.

Other potential applications for Wolter optics lie in inelastic scattering instruments, especially at the time-of-flight instruments where time-structure of the neutron beam is a factor. Other examples include diffraction from small samples and convergent-beam crystallography, where a converging beam on the sample is created by focusing mirrors [24]. In geometrical optics, ray paths can be reversed, therefore good focusing devices are usually also good collimators. Such optics could be used for collimating neutron beams, especially for the TOF instruments where preserving the time structure of the neutron pulse is important.

All of the applications outlined above require that the mirrors are coated with high-critical-angle multilayer coatings obtained by thin-film deposition techniques, as done for x-ray supermirrors [25]. Surface figure accuracy and roughness determine the angular resolution of grazing incidence optics. Good surface quality (roughness of 2 to 4 Å (rms)) is readily achievable for mirrors prepared by the replication technique. The angular resolution of one mirror pair is of 10 arc-seconds half power diameter (HPD), while nested systems are of 15 to 30 arc-seconds HPD. Therefore, the surface quality of our Wolter optics is at least similar, if not better, than of modern neutron guides. Specific requirements for the surface quality depend on applications. For instance, SANS requires low diffuse scattering from surface roughness, which is comparable to that of commercial neutron guides, while imaging requires small figure errors to achieve high resolution. A point source imaged by mirrors of 30 arc-seconds angular resolution and of 10 m focus-to-focus length will be of 0.2 mm HPD [10], while the image is magnified several-fold. Therefore, existing mirrors should be more than adequate for neutron imaging tasks, where detector resolution is of 0.1 mm.

## 6. Conclusions

We manufactured and tested neutron Wolter mirrors, and we demonstrated neutron beam focusing by a system of four nested Ni mirrors. Ray-tracing simulations of Wolter optics were carried out using McStas, incorporating both practical constraints of mirror fabrication and common geometrical layout of modern neutron instruments. Simulations showed that Wolter mirrors could be made to collect neutrons from sources as large as 10 mm diameter and demagnify the source by a factor of 10. It was found that for low-energy neutrons, the ratio between sample and source flux densities approach 10, provided both the sample and the source are in foci. Multilayer supermirror coating is extremely important for efficient utilization of neutron Wolter optics. The development of the coatings is currently under way.

Most importantly, we believe that there are several neutron techniques, which can benefit from the development of Wolter optics. The virtues of the Wolter design are low aberrations and axial symmetry [6]. First, the axial symmetry of the optics is very useful since it allows nesting of several mirror pairs inside each other to increase collection ability of the optics. Second, low aberrations mean that the optics creates a nearly true image of a sample. Furthermore, all neutrons reflected by one mirror pair travel the same amount of time between the source and focus. This feature is a major attribute for coupling with the new and powerful spallation sources.

## 7. Acknowledgments

Research supported by the U.S. Department of Energy, Office of Basic Energy Sciences, under Awards # DE-FG02-09ER46556 and DE-FG02-09ER46557 (Wolter optics studies) and by National Science Foundation under Award # DMR-0526754 (construction of Neutron optics test station and diffractometer at MIT).

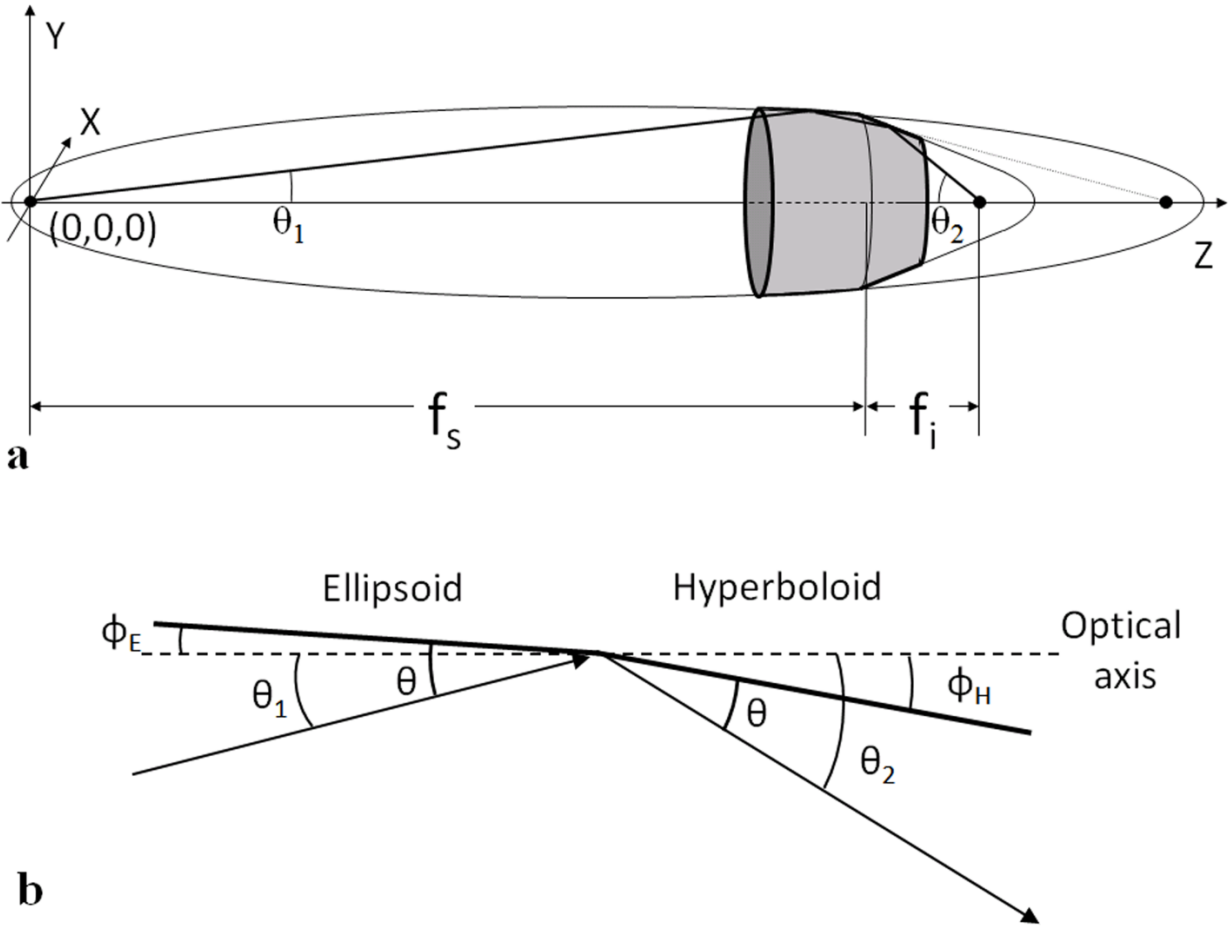

**Figure 1. (a)** Schematic view of a pair of Wolter focusing mirrors consisting of co-focal ellipsoid and hyperboloid. The small ("point") source is at the origin, coincident with the left focal point of the ellipsoid. The right focus (image) coincides with the left focal point of the hyperboloid. The right focal points of the ellipsoid and hyperboloid (the right-most dot on OZ axis) are coincident. The beam and optical axes coincide with OZ axis. The distance $f_s$ is between the source and intersection of the mirrors. The radius $r_i$ at intersection and the length of each mirror are input parameters. The distance between the intersection and the image is $f_i$, while $\theta_1$ and $\theta_2$ are the angles between incident and reflected rays and the optical axis OZ. **(b)** Schematic drawing of a cross-section of the mirrors (and rays) near intersection. The mirrors are shown in bold, the arrows are the rays and the dashed line is the optical axis. The angles $\theta_1$ and $\theta_2$ are between the rays and the optical axis, $\theta$ is between the rays and the mirrors, and $\varphi_E$ and $\varphi_H$ are between the tangent of the mirrors at the intersection point and the optical axis.

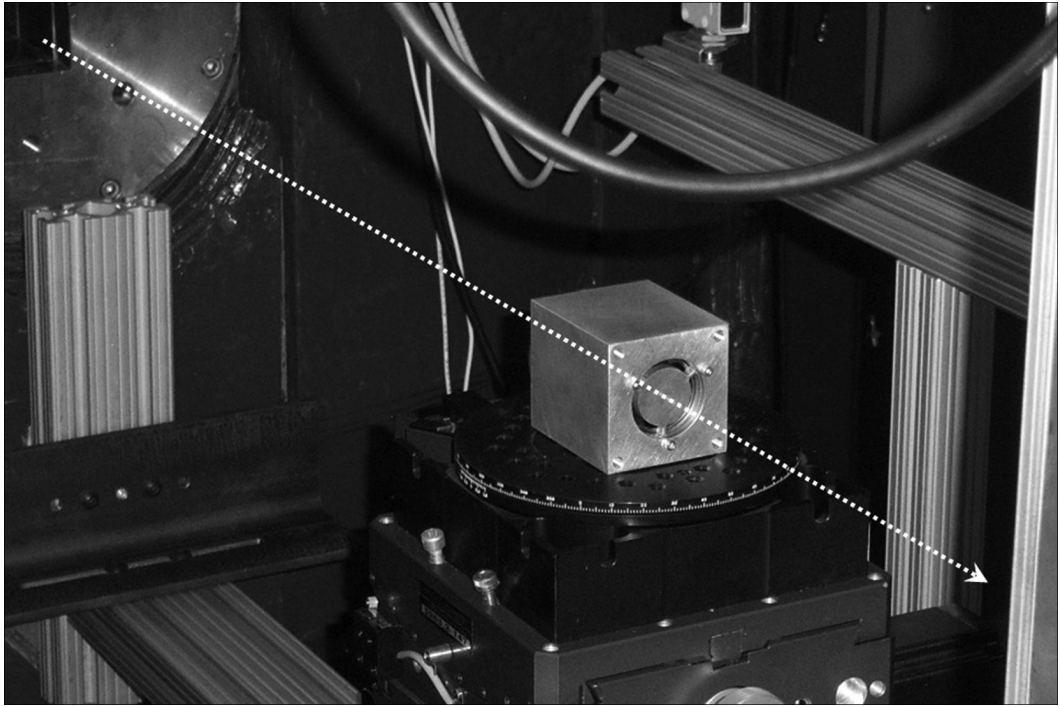

**Figure 2.** Wolter mirrors at the neutron beamline at the MIT Reactor. The neutron beam follows the dashed arrow from the top left to the bottom right corner of the picture. The mirrors inside an Al holder are positioned on top of a goniometer.

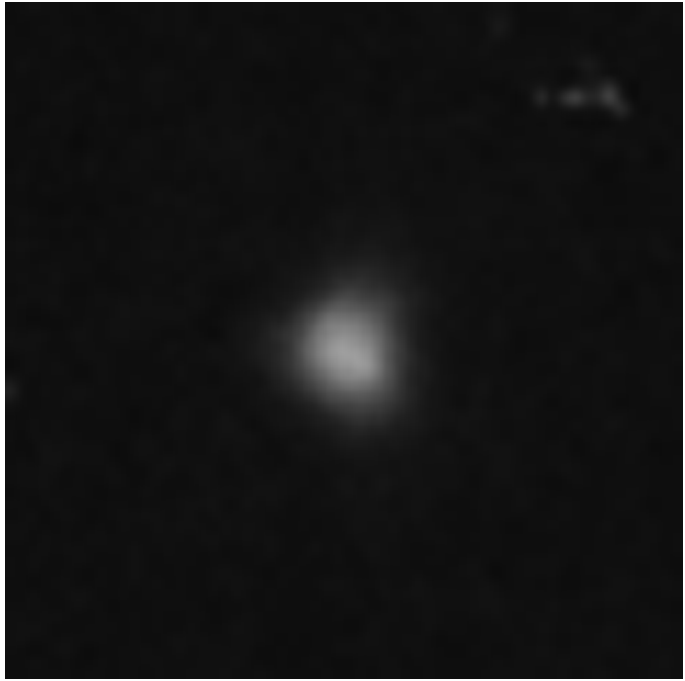

**Figure 3.** Image of the source when the detector is at the focal plane.

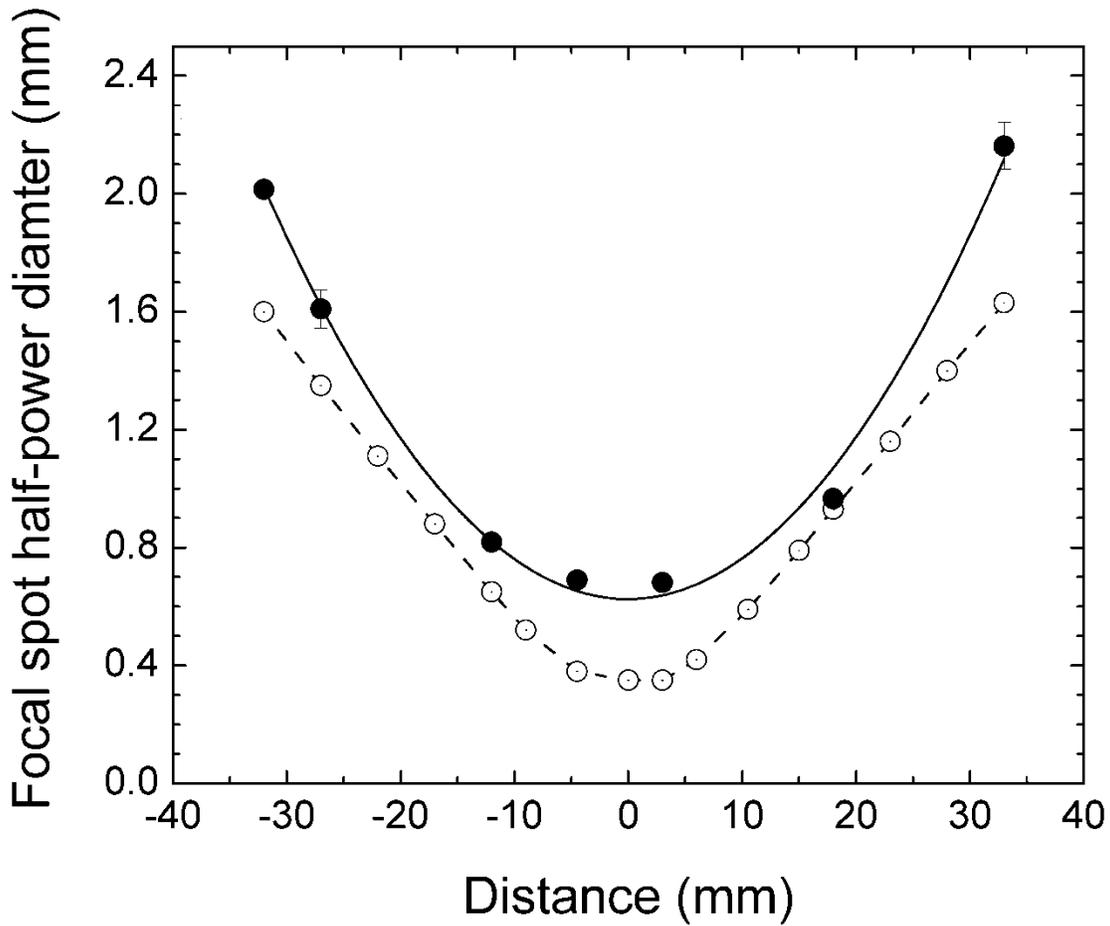

**Figure 4.** Depth of focus scan. The size of the focal spot was measured while the detector was scanning across the focal plane (positive distance is upstream of the focal plane). Large dots denote experimental measurements; circles connected by the broken line are ray-tracing calculations. Half-power diameter (HPD) was measured by fitting a Gaussian to the measured cross-section of the focal spot. The errors were deduced from the fit. Solid line is the parabolic fit of the experimental data. The fitted minimum of HPD is 0.62 mm. Calculated minimum of HPD is 0.35 mm. The discrepancy between the measurements and calculations is due to the misfit between the mirrors and their holder, as explained in the text.

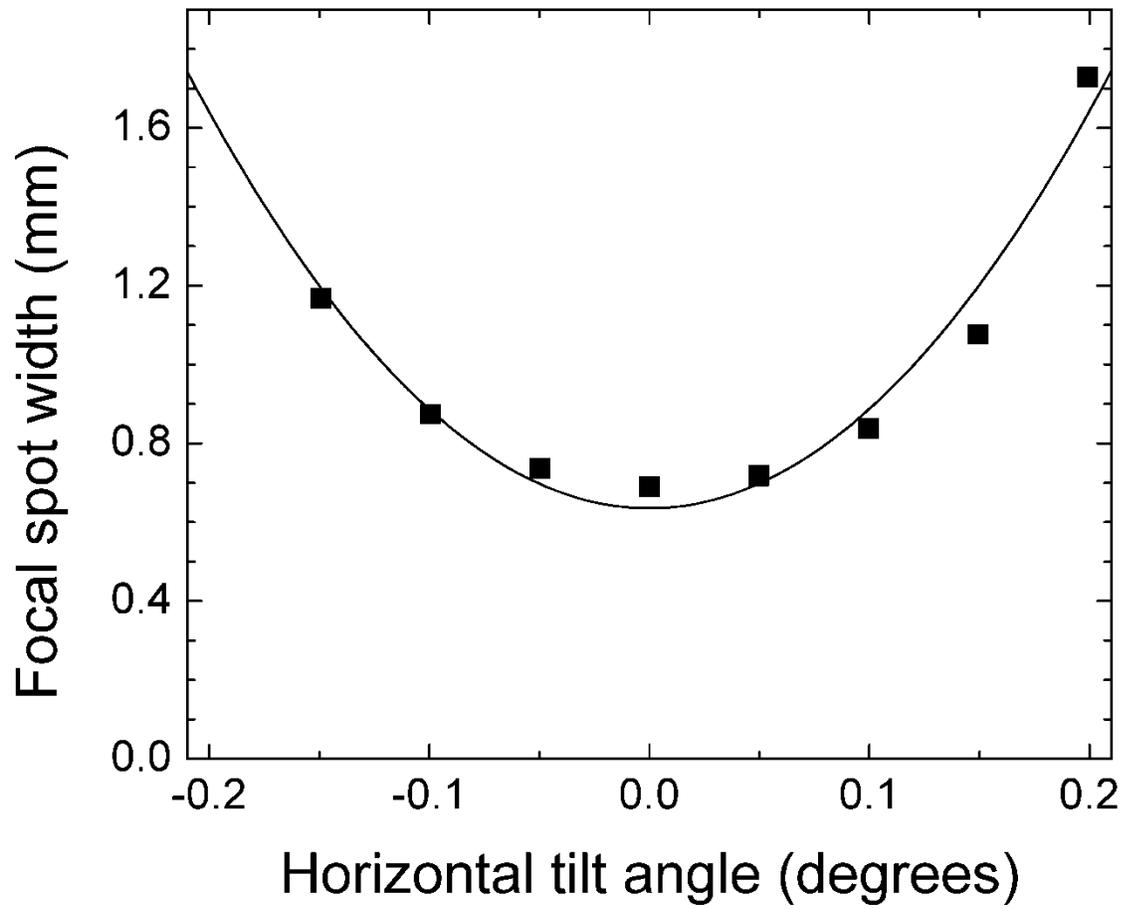

**Figure 5.** Experimental scan of the focal spot size vs mirrors' tilting angle in the horizontal plane. The line is a parabolic fit to the data.

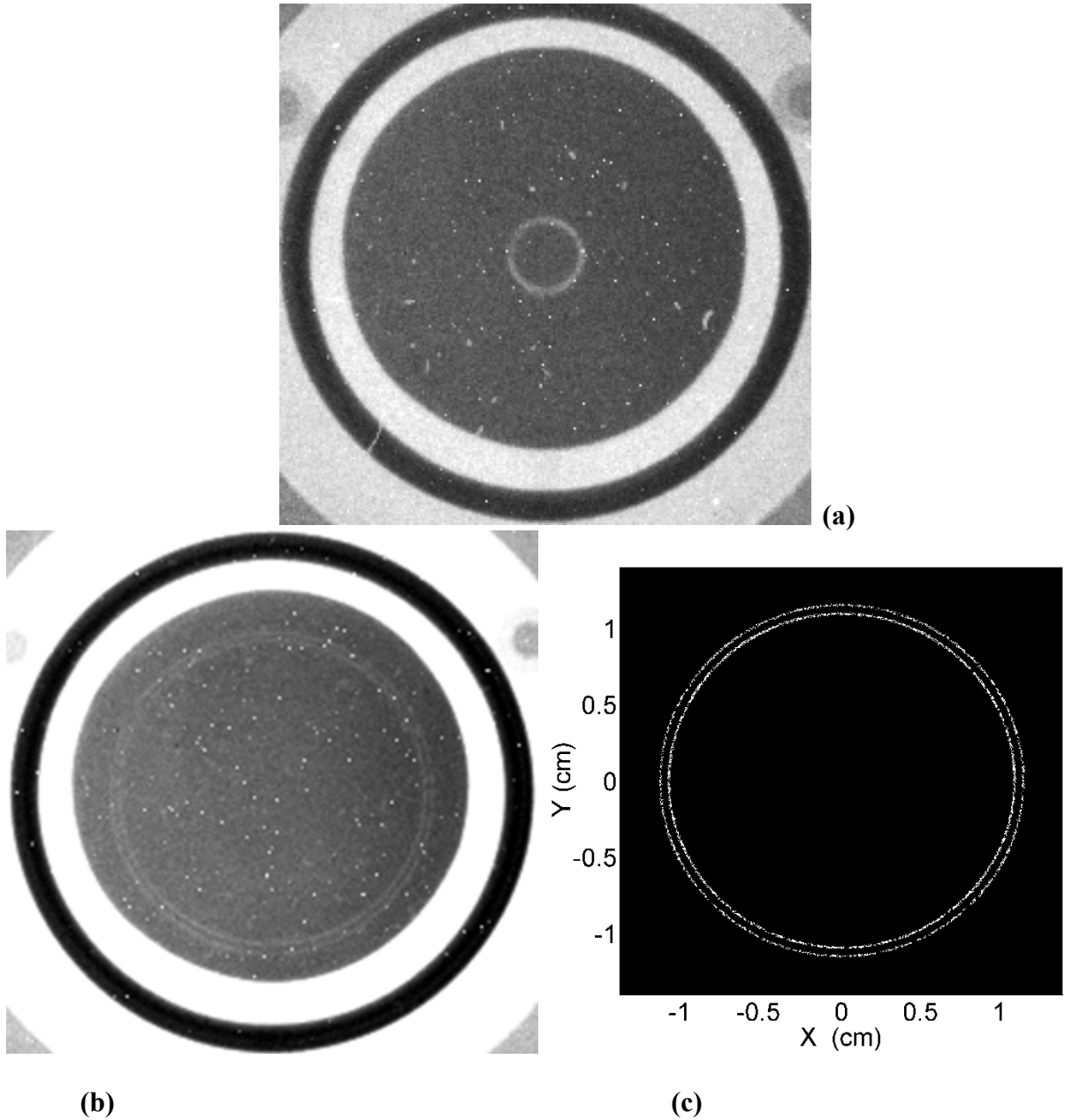

**Figure 6.** The images at the detector positioned behind a set-up of 2 nested mirrors, far away from the focal plane. (a) The distance between the mirrors intersection and the detector is 520 mm, while the focal plane distance is 640 mm from the intersection. The bright background is due to the neutron beam. The thick dark ring is the shadow of the mirrors. The gray circle inside the mirrors is the shadow of the Cd beam stop. The light ring in the center is due to the neutrons reflected by the mirrors. (b) The distance between the mirrors intersection and the detector is 200 mm. Two light, thin rings are visible inside the shadow of the beam-stop. These rings are created by the neutrons reflected by the two concentric mirrors. The neutrons in the two light rings converge onto the focal point. (c) Ray-tracing simulation of the geometry in the Figure 6(b).

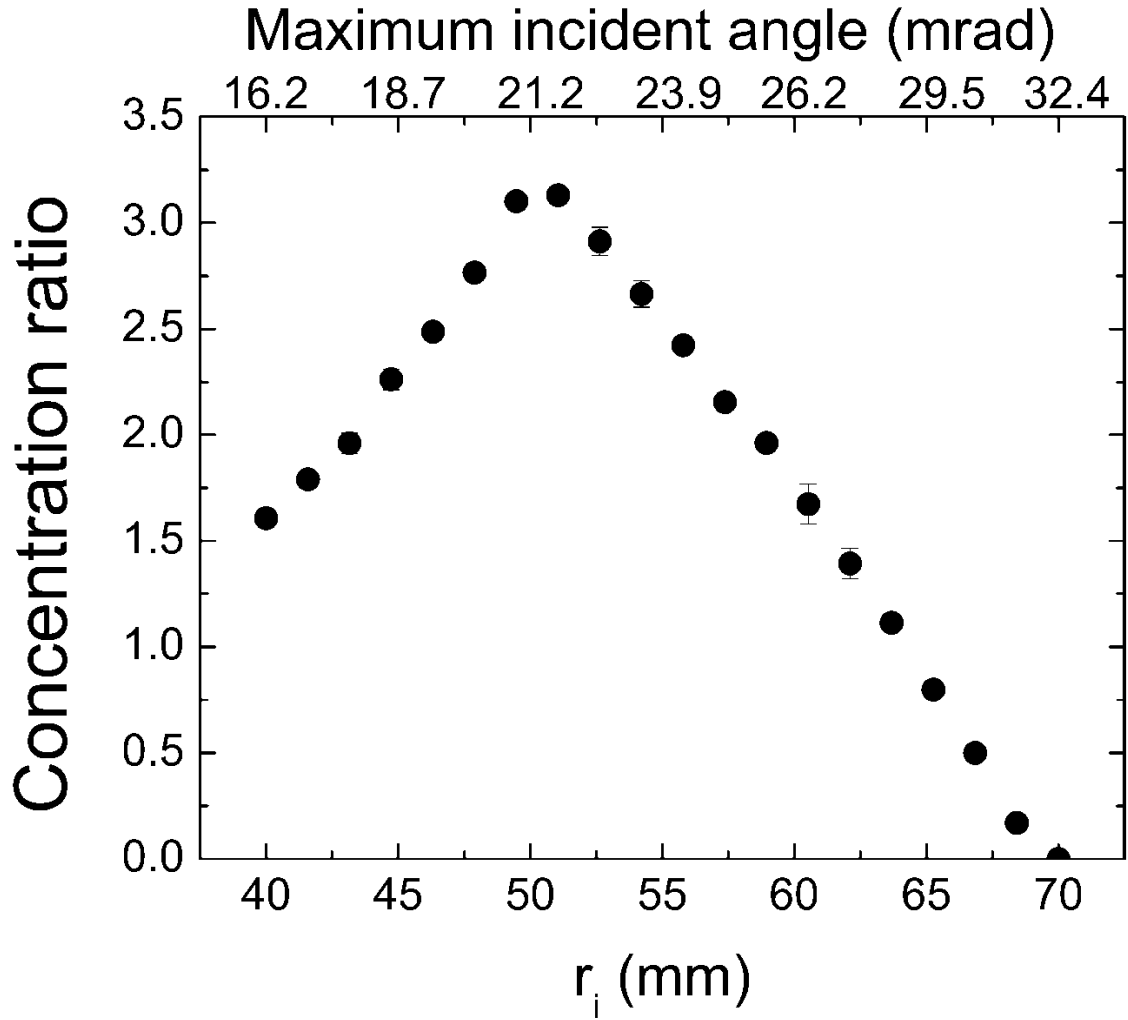

**Figure 7.** Ray-tracing simulations of large supermirror-coated Wolter optics described in Chapter 4 (m =3 supermirrors, E = 5 meV neutrons, source-to-focus distance 10 m, magnification M = 0.1). The concentration ratio, the ratio of flux densities at the detector and at the source, is calculated vs. the mirror's radius. The corresponding maximum grazing angle is plotted on the top axis.

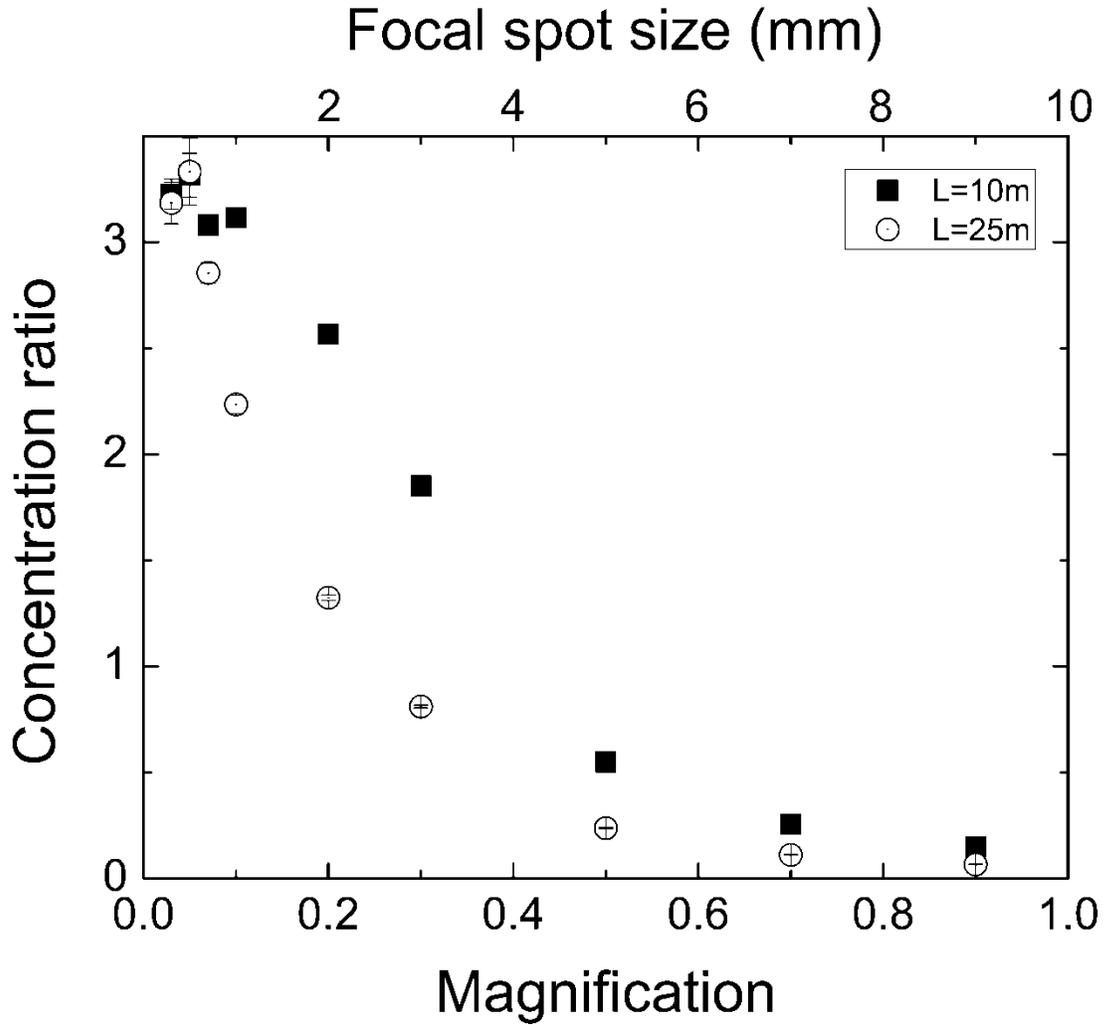

**Figure 8.** Flux concentration ratio as a function of magnification for two different source-to-sample distances: 10 m (squares) and 25 m (dots). Calculations are for E = 5 meV neutrons, m = 3 supermirror multilayer coating; each mirror is 0.7 m long. The size of the focal spot, shown on the top axis, is calculated assuming the source diameter is 10 mm.

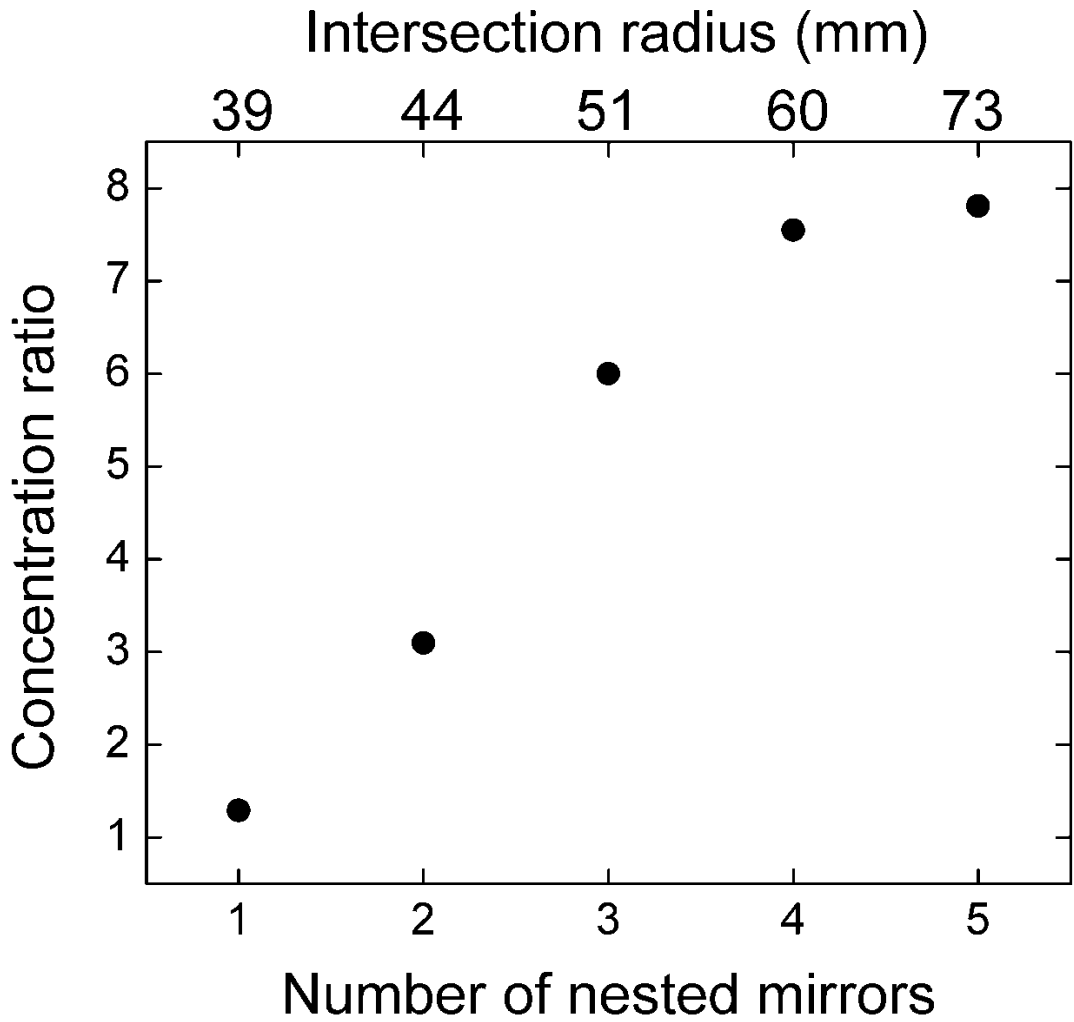

**Figure 9.** Concentration ratio of a system of several nested Wolter mirror pairs vs. the number of the mirror pairs. Calculations are for 5 meV neutrons, source to object distance of 10 m, m = 3 supermirror multilayer coating; each mirror is 0.7 m long.

# References

[1] P Böni. New concepts for neutron instrumentation, Nuclear Instruments and Methods in Physics Research Section A: Accelerators, Spectrometers, Detectors and Associated Equipment. 586 (2008) 1-8.

[2] GE Ice, CR Hubbard, BC Larson, JW L. Pang, JD Budai, S Spooner, et al. Kirkpatrick–Baez microfocusing optics for thermal neutrons, Nucl. Instr. Meth. A. 539 (2005) 312-320.

[3] G Ice, C Hubbard, B Larson, J Pang, J Budai, S Spooner, et al. High-performance Kirkpatrick-Baez supermirrors for neutron milli-and micro-beams, Materials Science & Engineering A. 437 (2006) 120-125.

[4] AD Stoica, XL Wang, WT Lee, JW Richardson. Multiple-stage tapered neutron guide as a broad band focusing system, Proc. SPIE. 5536 (2004) 86.

[5] M Gubarev, B Ramsey, D Engelhaupt, J Burgess, D Mildner. An evaluation of grazing-incidence optics for neutron imaging, Nuclear Inst.and Methods in Physics Research, B. 265 (2007) 626-630.

[6] H Wolter. Spiegelsysteme streifenden Einfalls als abbildende Optiken fur Rontgenstrahlen, Ann. Der Physik. 10 (1952) 52.

[7] L VanSpeybroeck, R Chase. Design parameters of paraboloid-hyperboloid telescopes for x-ray astronomy, Appl.Opt. 11 (1972) 440-445.

[8] BD Ramsey, RF Elsner, D Engelhaupt, MV Gubarev, JJ Kolodziejczak, SL ODell, et al. The development of hard X-ray optics at MSFC, Proc. of SPIE. 4851 (2003) 631-638.

[9] BD Ramsey. Replicated Nickel Optics for the Hard-X-Ray Region, Exp. Astron. 20 (2005) 85-92.

[10] M Pivovaroff, T Funk, W Barber, B Ramsey, B Hasegawa. Progress of focusing x-ray and gamma-ray optics for small animal imaging, Proc. SPIE. 5923 (2005) 59230B.

[11] K Lefmann, K Nielsen. McStas, a general software package for neutron ray-tracing simulations, Neutron News. 10 (1999) 20-23.

[12] P Willendrup, E Farhi, K Lefmann. McStas 1.7-a new version of the flexible Monte Carlo neutron scattering package, Physica B. 350 (2004) E735-E737.

[13] R Winston, JC Minano, W Welford, P Benítez, Nonimaging optics, Academic Press, Amsterdam ; Boston, Mass., 2005.

[14] B Aschenbach. X-Ray Telescopes, Rep. Prog. Phys. 48 (1985) 579-629.

[15] AG Michette. X-ray microscopy, Rep. Prog. Phys. 51 (1988) 1525-1606.

[16] KD Joensen, P Gorenstein, O Citterio, P Hoghoj, IS Anderson, O Schaerpf. Hard x-ray Wolter-I telescope using broadband multilayer coatings on replica substrates: problems and solutions, Proc. of SPIE. 2515 (1995) 146.

[17] B Alefeld, L Dohmen, D Richter, T Brückel. X-ray space technology for focusing small-angle neutron scattering and neutron reflectometry, Physica B: Condensed Matter. 283 (2000) 330-332.

[18] D Mildner. Resolution of small-angle neutron scattering with a refractive focusing optic, Journal of Applied Crystallography. 38 (2005) 488-492.

[19] T Oku, T Shinohara, J Suzuki, R Pynn, HM Shimizu. Pulsed neutron beam control using a magnetic multiplet lens, Nuclear Instruments and Methods in Physics Research Section A: Accelerators, Spectrometers, Detectors and Associated Equipment. 600 (2009) 100-102.

[20] T Shinohara, S Takata, J Suzuki, T Oku, K Suzuya, K Aizawa, et al. Design and performance analyses of the new time-of-flight smaller-angle neutron scattering instrument at J-PARC, Nuclear Instruments and Methods in Physics Research Section A: Accelerators, Spectrometers, Detectors and Associated Equipment. 600 (2009) 111-113.

[21] H Beguiristain, I Anderson, C Dewhurst, M Piestrup, J Cremer, R Pantell. A simple neutron microscope using a compound refractive lens, Appl.Phys.Lett. 81 (2002) 4290.

[22] BE Allman, PJ McMahon, KA Nugent, D Paganin, DL Jacobson, M Arif, et al. Phase radiography with neutrons, Nature. 408 (2000) 158-159.

[23] N Kardjilov, E Lehmann, E Steichele, P Vontobel. Phase-contrast radiography with a polychromatic neutron beam, Nucl. Instr. Meth. A. 527 (2004) 519-530.

[24] W Gibson, A Schultz, J Richardson, J Carpenter, D Mildner, H Chen-Mayer, et al. Convergent-beam neutron crystallography, Journal of Applied Crystallography. 37 (2004) 778-785.

[25] S Romaine, J Boike, R Bruni, D Engelhaupt, P Gorenstein, M Gubarev, et al. Mandrel replication for hard x-ray optics using titanium nitride, Proceedings of SPIE. 7437 (2009) 74370Y.